\documentclass[preprint,preprintnumbers,amsmath,amssymb]{revtex4-1}
\usepackage{graphicx}
\begin{document}
%\begin{CJK}[HL]{KS}{}
%=======================================================================================================
\title[Short Title]{The way to uncover community structure with core and diversity}
\author{\bf YunFeng Chang$^{1,2}$\footnote{Corresponding author E-mail: changyf@ctgu.edu.cn\\
$^{1}$College of Science, China Three Gorges University, Yichang 443002, Hubei, P. R. China\\
$^2$ Department of Physics, Chungbuk National University, Cheongju 28644, Chungbuk, R. Korea}, SeungKee Han$^{2}$, XiDong Wang$^{1}$}
%\affiliation{$^1$ College of Science, China Three Gorges University, P. R. China\\
%$^2$ Department of Physics, Chungbuk National University, R. Korea}
%=======================================================================================================
\begin{abstract}
Communities are ubiquitous in nature and society. Individuals that share common properties often self-organize to form communities. Being able to identify community structure could help us understand and explore complex systems efficiently. Avoiding the shortages of computation complexity, pre-given information and unstable results in different run, in this paper, we propose one simple and efficient method to try to give a deep understanding of the emergence and diversity of communities in complex systems. By introducing rational random selection, our method reveals the hidden deterministic and normal diverse community states of community structure. To demonstrate this method, we test it with real-world systems. The results show that our method could not only detect community structure with high sensitivity and reliability, but also could provide instructional information about our normal diverse community world and  the hidden deterministic community world by giving out the core-community, the real-community, the tide (boundary) and the diversity. This is of paramount importance in understanding, predicting, and controlling a variety of collective behaviors in complex systems.\par
%{\bf PACS number:}\ 75.60.Ch,\ 75.40.Gb,\ 75.40.Mg\par
\end{abstract}\maketitle
%=======================================================================================================
{\bf Introduction}

Communities are supposed to play special roles in the structure-function relationship. For examples, the communities in WWW are sets of web pages sharing the same topic\cite{1}; the modular structure in biological networks are widely believed to play important roles in biological functions\cite{10,2,3,4,5}. The identification of community structure helps when analyzing the functionalities and organizations of complex systems.

With the spring up of complex network, which have attracted considerable attention in physics and other fields as a foundation for the mathematical representation of a variety of complex systems, many systems in different areas such as biology\cite{3}, sociology\cite{6}, medicine\cite{7}, web\cite{8}, and many others\cite{9} are represented as networks. In the field of complex network study, communities are defined as groups of nodes that are densely interconnected but only sparely connected with the rest of the network\cite{10,11,42,43}. With this network based definition, researchers have proposed different algorithms for detecting communities such as topology based methods\cite{10,33}, modularity optimization\cite{34,35}, dynamic label propagation\cite{36,37,38}, statistical inference\cite{39,40,41}. Besides the shortages of computation complexity, pre-given information and unstable results in different run, moreover, some social networks are found with different community property that the individuals in one group might be gregarious, having many contacts with others, while the individuals in another group might be more reticent. An example of this behavior is seen in networks of sexual contacts, where separate communities of high- and low-activity individuals have been observed\cite{17,18}. That is why there is no commonly agreed definition for community. For reviews see reference\cite{26,27}

Community detection is also called cluster analysis which is done with different kinds of relationships. Specifically, cluster analysis is the assignment of a set of observations into clusters of components that are similar to each other but different from components in other clusters. It is often used to ascertain whether a complex system comprises a set of distinct clusters, each representing components with substantially different properties. The segmentation of complex systems into clusters could also allow us to find specific functions naturally assigned to each cluster, as in the case of human functional brain system\cite{19} and metabolic system\cite{3}. A number of clustering methods have been developed as a tool for handling large and heterogeneous collections of systems, e.g., hierarchical clustering, k-means clustering, and affinity propagation\cite{20,21,22}.

Incorporating the two community detection considerations above, the elusive question is how communities with common internal properties arise? Inspired by the ultimatum game experiment in reference\cite{23}, in this paper, we propose one simple and efficient community detection method based on this elusive question. This method is able to cluster complex systems efficiently by merging components of the closest proximity into the same community. During the merging process, characteristic numbers of communities are obtained for the complex systems, which correspond to various resolution scales for viewing the system. Moreover, we get two practical community states, hidden deterministic and normal diverse, by introducing rational random selection. These two states provide an in-depth view of the community structure of real-life complex systems with diversity. To demonstrate this method, we test it with real-world systems and find that the method could not only detect community structure with high sensitivity and reliability but also could provide us instructional information about our community world by giving out the core-community, the real-community, the tide (boundary) and the diversity. This is of paramount importance in understanding, predicting, and controlling a variety of collective behaviors in complex systems, especially social complex systems.

{\bf Method}

For finding meaningful communities, it is better to follow the real process of community formation or construction. Han et al. tried to explore this process by ultimatum game experiment\cite{23}. In real-life systems, communities are constructed by individuals with the choosing of friends. And this choosing process are base on individual's judgment of its relationship with the surroundings. Most probably, individual chooses the one who is most similar to it or the one satisfy its expectation mostly. So that it might be a good choice to reconstruct and detect communities by the most similar pairs. That means, detecting communities by formalizing those relationships or those components believed to be the most significant.

The procedure of our method is explained in the following steps using Zachary's karate club\cite{25} as an example. The constructed Zachary's karate club, a university-based karate club, consists of 34 nodes. At the beginning of Zachary's study there was an incipient conflict between the club instructor ({\bf node 1}) and administrator ({\bf node 34}) over the price of karate lessons. As time passed, the entire club became divided over this issue during the course of Zachary's study. The graph representation of the relationships in the club (shortly before the fission) can be seen in \cite{25}. The karate system is represented as a graph $G(V,E)$ with a set of nodes $V=v_{1},v_{2},...,v_{34}$ model the members of the club and a set of edges  indicating that two individuals consistently were observed to interact outside the normal activities of the club. 

Similarity can be different kinds of interaction according to the properties of complex systems. For examples, internet users with common interests\cite{28}, social communities with distinctive social norms form spontaneously\cite{29}, related proteins group together to execute specific functions within a cell\cite{30}. Here, for the karate club, connected common neighbor is a good choice because human has the intention to follow the major, this is also the basic idea of label propagation algorithm. Like a friendship network, each person affects their known people. More common known people make two acquaintances more close to each another. The neighborhood of node $u\in V$ is the set of adjacent nodes of $u$,

\begin{equation}
\Gamma_{u}=\{v\in V|(u,v)\in E\}
\end{equation}

Common neighbors of two nodes $u,v\in V$ is the set of nodes $\Gamma_{uv}$ containing adjacent nodes of node $u$ that are also adjacent nodes of node $v$,

\begin{equation}
\Gamma_{uv}=\{p\in V|(u,p)\in E \& (v,p)\in E\}
\end{equation}

With the adjacency matrix $A$, where $A_{ij}=1$ if nodes $i$ and $j$ are connected and  otherwise $A_{ij}=0$. Connected common neighbor (similarity) can be calculated in the following way:

\begin{equation}
S_{uv}=A_{uv}+\Gamma_{uv}=A_{uv}(1+\sum\limits_{k} A_{uk}*A_{kv})
\end{equation}

With $S_{uv}$, our method is carried out in the following steps:

Step 1: Identify all the most similar neighbor of each node and record their similarity in decreasing order, as shown in Table 1. For a system with $N$ components, our method needs only to process the $O(N)$ elements in this list instead of dealing with a matrix of $N^2$ elements.

\begin{table}[!htp]
\begin{center}
\begin{tabular}{|c c c | c c c |c c c|c c c | c c c | c c c |}
\hline
$u$ & $v$ & $S_{uv}$ & $u$ & $v$ & $S_{uv}$ & $u$ & $v$ & $S_{uv}$ & $u$ & $v$ & $S_{uv}$ & $u$ & $v$ & $S_{uv}$ & $u$ & $v$ & $S_{uv}$ \\
\hline
33 & 34 & 11    & 9 & 33 & 4      & 7 & 1 & 3 &       15 & 34 & 2 &       20 & 2 & 2         & 26 & 32 & 2\\
34 & 33 & 11    & 14 & 1 & 4        & 7 & 6 & 3 &       16 & 33 & 2 &     21 & 33 & 2          & 27 & 30 & 2\\
1 & 2 & 8          &14 & 2 & 4 &      11 & 1 & 3 &          16 & 34 & 2       & 21 & 34 & 2          & 27 & 34 & 2 \\
2 & 1 & 8          & 14 & 3 & 4 &       31 & 9 & 3 &         17 & 6 & 2        & 22 & 1 & 2        & 28 & 24 & 2\\
3 & 1 & 6          & 14 & 4 & 4 &      31 & 33 & 3 &         17 & 7 & 2         & 22 & 2 & 2      & 28 & 34 & 2\\
4 & 1 & 6          & 24 & 34 & 4 &     31 & 34 & 3 &         18 & 1 & 2         & 23 & 33 & 2       & 29 & 32 & 2\\
8 & 1 & 4         & 30 & 34 & 4 &        32 & 34 & 3 &        18 & 2 & 2        & 23 & 24 & 2       & 29 & 34 & 2\\
8 & 2 & 4           & 5 & 1 & 3 &          13 & 1 & 2 &            19 & 33 & 2        & 25 & 26 & 2      & 10 & 3 & 1\\
8 & 3 & 4          & 6 & 1 & 3 &          13 & 4 & 2 &          19 & 34 & 2          & 25 & 32 & 2      & 10 & 34 & 1\\
8 & 4 & 4         & 6 & 7 & 3 &           15 & 33 & 2 &          20 & 1 & 2         & 26 & 25 & 2      & 12 & 1 & 1\\
\hline
\end{tabular}
\renewcommand{\thetable}{\arabic{table}}
\renewcommand{\tablename}{Table}
\caption{Selected most similar node pairs according to connected common neighbor}
\end{center}
\label{default}
\end{table}%

Step 2: Communities are constructed by starting from the connected node pair with the maximum similarity and then including more connected node pairs from the list. In this example, to begin with, node 33 and 34 formed the first core-community $CCom_{1}=\{33,34\}$. Then node 1 and node 2 formed the second core-community $CCom_{2}=\{1,2\}$. $CCom_{2}$ grew by including nodes 3, 4 through their connections with node 1 and resulted in $CCom_{2}=\{1,2,3,4\}$. Node 8 is included into $CCom_{2}$ through any one of its connections with nodes 1,2,3,4 resulted in $CCom_{2}=\{1,2,3,4,8\}$. Similarly, node 9 is added to $CCom_{1}$, node 14 is added to $CCom_{2}$, node 24 and 30 is added to $CCom_{1}$, node 5, 6, 7, 11 to $CCom_{2}$; nodes 31, 32 to $CCom_{1}$, node 13 to the second, node 15, 16 to the first, node 17, 18 to the second, node 19 to the first, node 20 to the second, node 21 to the first, 22 to second, 23 to the first. Node 25 and 26 to $CCom_{1}$ through (25,32) and (26,32). Then nodes 27, 28, 29 are added to $CCom_{1}$. Node 10 is added to $CCom_{2}$. And $CCom_{1}$ and $CCom_{2}$ are connected by the tide (10,34). Finally, node 12 is added to $CCom_{2}$ through its connection to node 1.

$CCom_{1}=\{33,34,9,24,30,31,32,15,16,19,21,23,25,26,27,28,29\}$

$CCom_{2}=\{1,2,3,4,8,14,5,6,7,11,13,17,18,20,22,10,12\}$
%\end{equation}

%\begin{equation}
$RCom=\{CCom_{1},CCom_{2}\}$
%\end{equation}

The tide (10,34) causes the merging of $CCom_{1}$ and $CCom_{2}$ into the real-community $RCom$.

{\bf Results and Analysis}

{\bf The hidden deterministic state.} We call this community detection process with all the most similarity pairs: the hidden deterministic state. This is an ideal but usually impossible world for large complex systems because of the limitation of information and exact description. However, the hidden deterministic world does provide us with instructional community information about the system. Here, for the karate club, we could get the following instructions:

\begin{itemize}
\item The club has two hidden major core-communities with (33,34) and (1,2) acted as the central nodes. Detailed statistics gives out that, the core nodes are 34 and 1 because more nodes are connected to $CCom_{1}$ and $CCom_{2}$ through their connection to 34 and 1 correspondingly. This is in accordance with the real situation where node 34 is the administrator and node 1 is the instructor\cite{25}.

\item In the above community detecting process, all the most similar pairs for one individual are processed together. However, if we observe these node pairs in detail, more community information of the system can be discovered. Some of the nodes with two or more most similar node pairs have no obscurity in the selection of community, just like nodes 8, 14, 31, 13. However, there are also some special nodes deserve more attention. For example, node pairs for node 6. If we deal with (6,7) firstly, a new core-community $\{6,7\}$ will appear temporarily. As for the node pairs for node 25, a new core-community $\{25,26\}$ will be found. All these detailed community information has heuristic and instructional meaning in community structure analysis. We will explain them later in the part of normal diverse community world.

\item One vacillate node 10 is found, which has two most similar nodes 3 and 34 since it straddles between $CCom_{1}$ and $CCom_{2}$. That is because of the oversimplified relationship of the network data. By investigating the system more in detail, Zachary gives out that node 10 is more similar with node 34\cite{25}. So there is no doubt that node 10 belongs to $CCom_{1}$ leaded by the administrator node 34. It is also natural for node 10 choose $CCom_{1}$ because it is connected directly to the leader of $CCom_{1}$. However, the choice of node 10 is unknown before it made the choice, we can only guess that according to his behavior with probability, and made the conclusion that it may choose $CCom_{1}$ most probably.

\item Finally and importantly, further detection could be done for the hidden deterministic world by regarding the detected communities as coarse-grained components. Then the coarse-grained system comprising these renormalized components can be further classified by steps 1 and 2\cite{24}.
\end{itemize}

The hidden deterministic state shows that Zachary's karate club is a system with two hidden core-communities, that it will evolve into two parts leaded by the instructor node 1 and the administrator node 34.

{\bf The normal diverse state.} Then, is the above our real-life community world? In fact, it is more common and universal for individuals to choose one neighbor at one time according to their preference: rational random selection. We call the community detection with rational random selection the normal diverse state. The organization of real-life community world is usually based on such a rational random selection. 

Clearly, the hidden deterministic world would be the same with the normal diverse world if all the individuals have only one distinct most similar node pair. For the karate system, as showed in table 1, most nodes have two or more most similar node pairs, which results in a diverse community world. Table 2 is the detailed community detection results in the normal diverse state for the karate system in 10000 run. The normal diverse state shows that:

\begin{figure}[!htp]
\centering
\includegraphics[width=4cm,height=3cm]{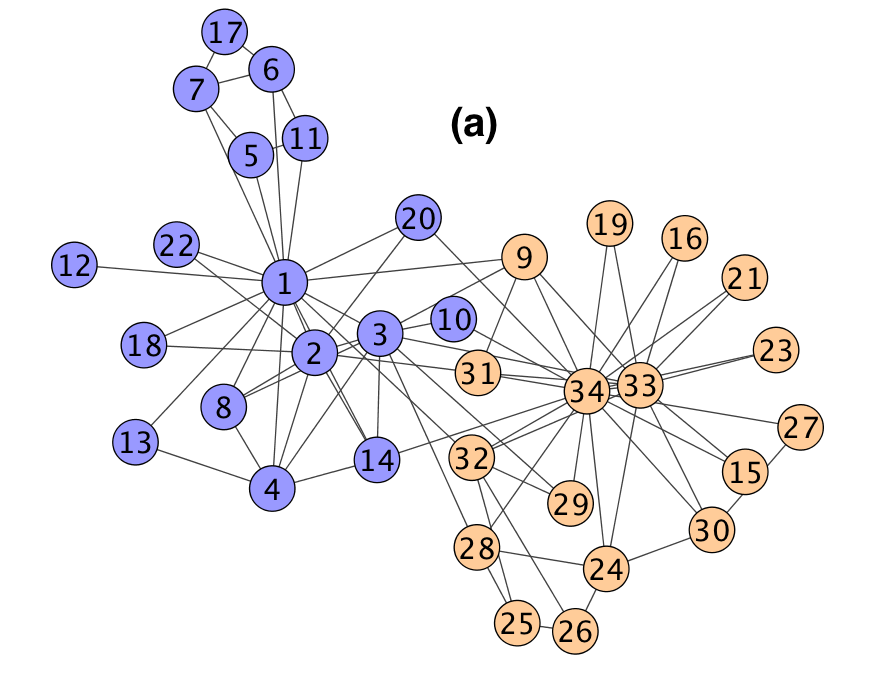}
%\hspace{0.5in}
\includegraphics[width=4cm,height=3cm]{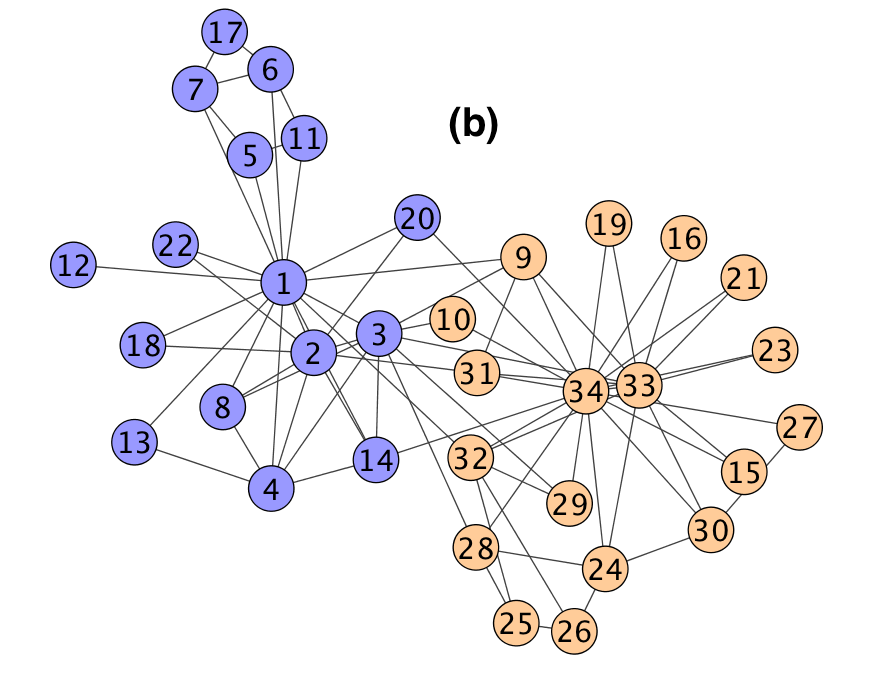}
\includegraphics[width=4cm,height=3cm]{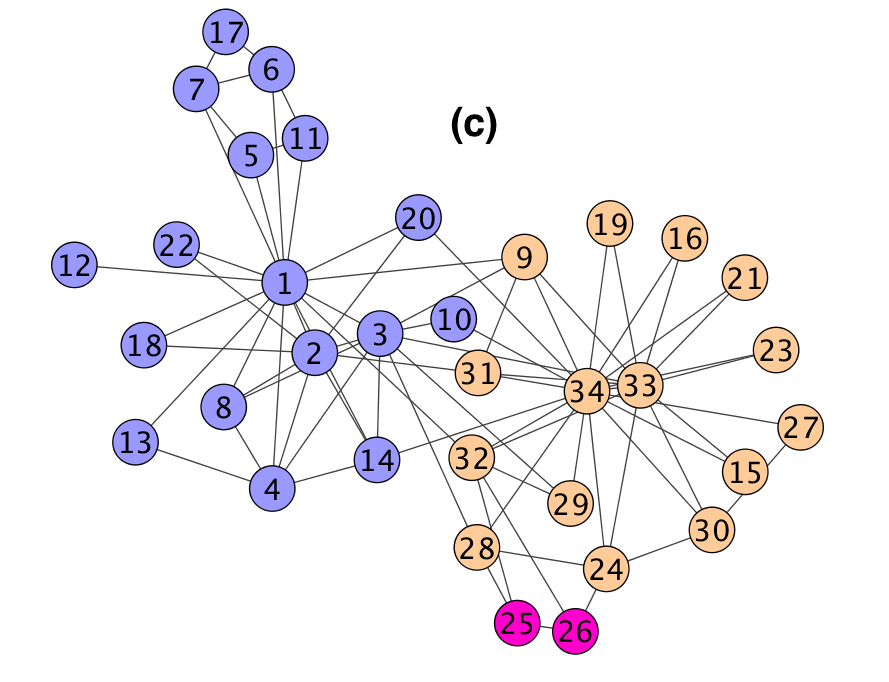}
%\hspace{0.5in}
\includegraphics[width=4cm,height=3cm]{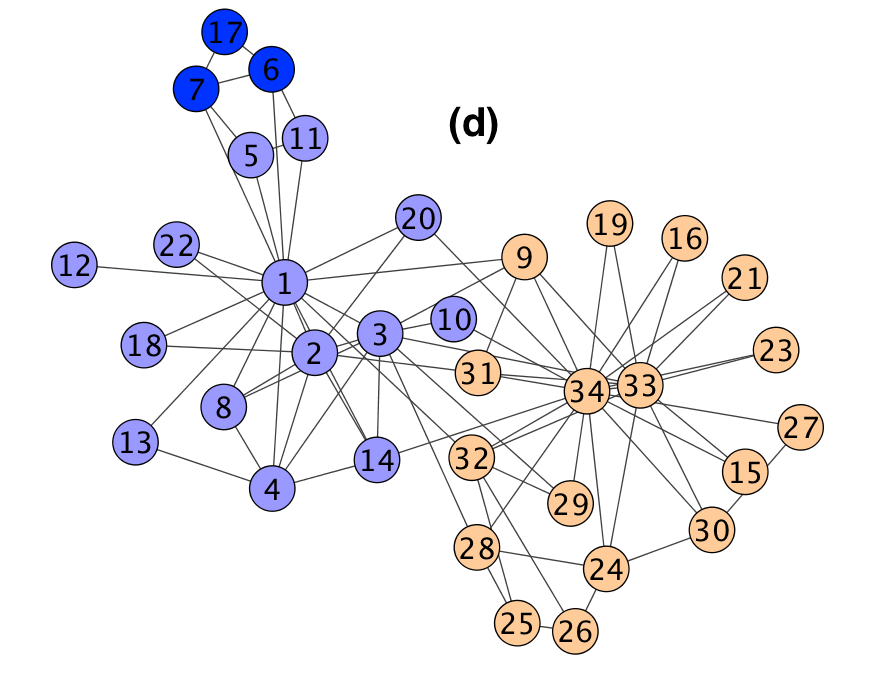}
\includegraphics[width=4cm,height=3cm]{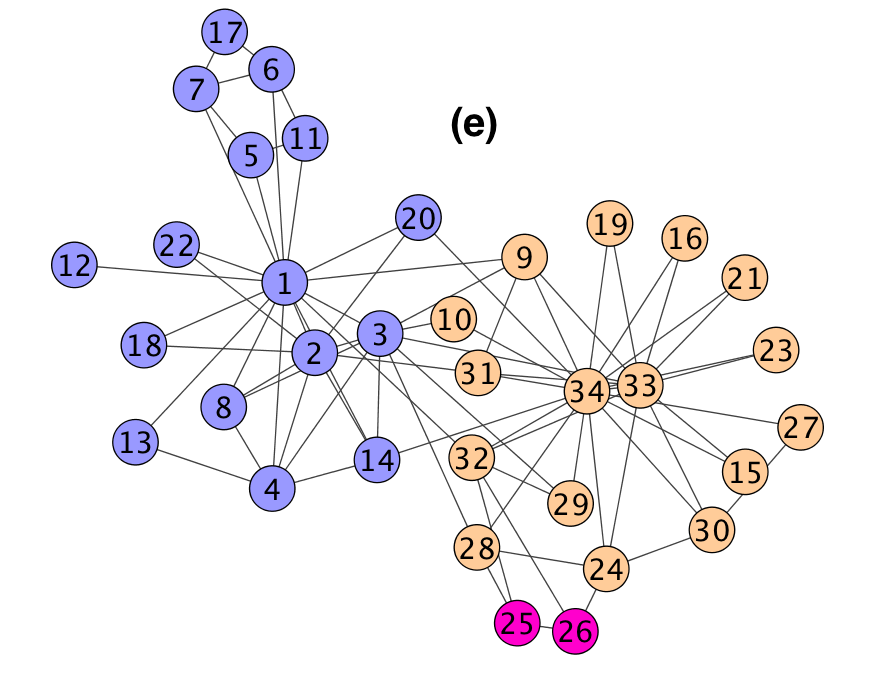}
%\hspace{0.5in}
\includegraphics[width=4cm,height=3cm]{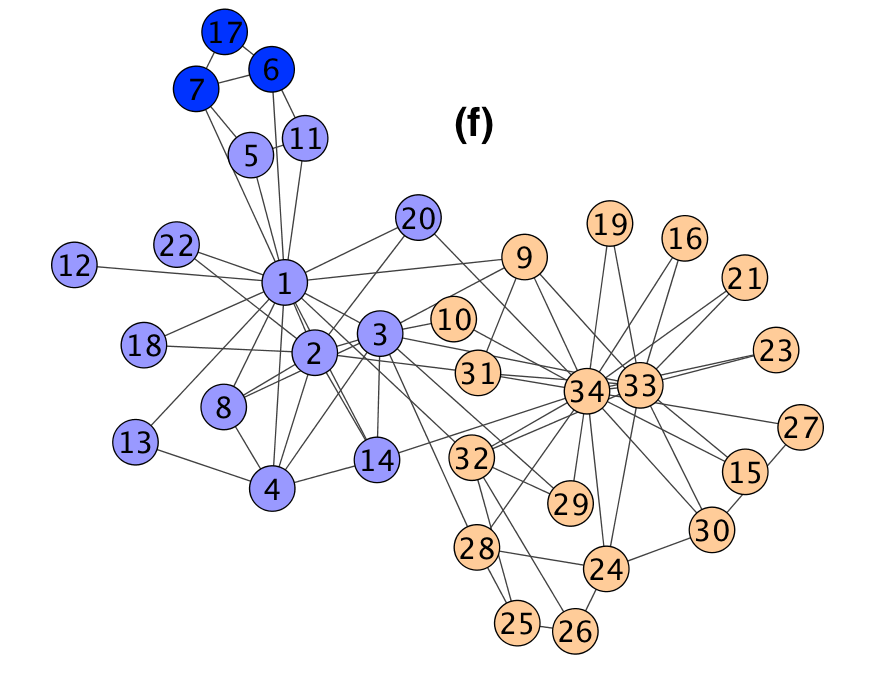}
\includegraphics[width=4cm,height=3cm]{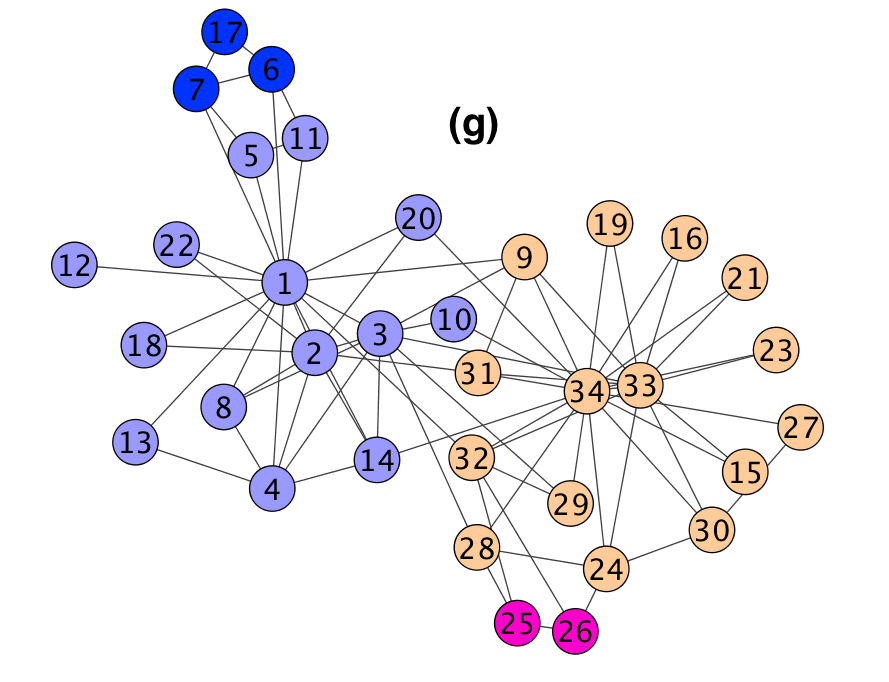}
%\hspace{0.5in}
\includegraphics[width=4cm,height=3cm]{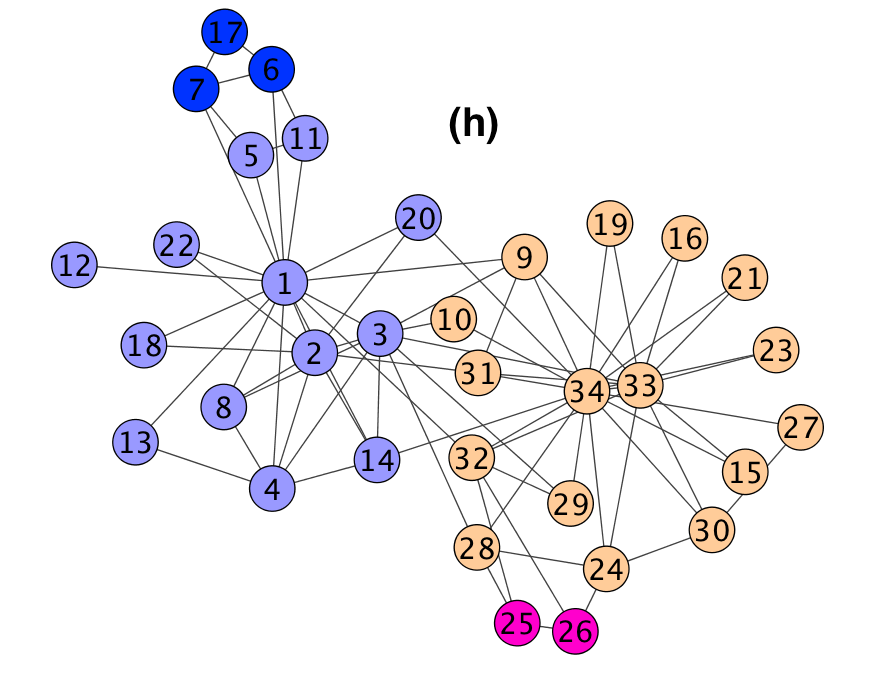}
\renewcommand{\figurename}{Figure}
\caption{All the detected normal diverse community structure of the karate system.}
\end{figure}

\begin{table}[!htp]
\begin{center}
\begin{tabular}{|c|c|c|c|c|}  
    \hline  
    & number of communities & number of appearance & detailed structure & number of appearance\\
    \hline
   & 2  & 2513 & Fig.1(a)(17,17) & 1261\\
    \cline{4-5}
  &&& Fig.1(b)(16,18) & 1252 \\
   \cline{2-5}
  CCom && & Fig.1(c)(15,17,2) & 1196\\
   \cline{4-5}
   &  3 &  4987  & Fig.1(d)(14,17,3) & 1255\\
   \cline{4-5}
  &&&  Fig.1(e)(16,16,2) & 1251\\
  \cline{4-5}
  && & Fig.1(f)(13,18,3) & 1285\\
   \cline{2-5}
  & 4 & 2500  & Fig.1(g)(14,15,2,3) & 1221\\
  \cline{4-5}
  &&& Fig.1(h)(13,16,2,3) & 1279\\
    
    \hline
   &  2  & 5634 & Fig.1(a)(17,17) & 2821\\
    \cline{4-5}
  &&&Fig.1(b)(16,18) & 2813 \\
   \cline{2-5}
    RCom & & & Fig.1(c)(15,17,2) & 906\\
   \cline{4-5}
   &  3  & 3733 & Fig.1(d)(14,17,3) & 909\\
   \cline{4-5}
  &&&  Fig.1(e)(16,16,2) & 945\\
  \cline{4-5}
  && & Fig.1(f)(13,18,3) & 973\\
   \cline{2-5}
  & 4 & 633   & Fig.1(g)(14,15,2,3) & 297\\
  \cline{4-5}
  &&& Fig.1(h)(13,16,2,3) & 336\\
 \hline 
\end{tabular}
\renewcommand{\thetable}{\arabic{table}}
\renewcommand{\tablename}{Table}
\caption{The normal diverse community detection results and the number of its appearance in 10000 run. The attached numbers in detailed structure are the number of members in different community.}
\end{center}
\label{default}
\end{table}%

\begin{itemize}
\item After introducing rational random selection, stable small local communities ($\{25,26\}$ and $\{6,7,17\}$) emerge spontaneously and induce social diversity into the system, this phenomenon is also found experimentally by Han et. al. in \cite{23}.  $\{25,26\}$ is recognized as a small community because both of them settle at the outer boundary of $CCom_{2}$, as well as, with no connection with $CCom_{1}$. One can get this fact from Figure 1. For the same reason, $\{6,7,17\}$ is recognized as a small community, and it is included into $CCom_{1}$ in the fission because it sides on the outer boundary of $CCom_{1}$ while it has no connection with $CCom_{2}$. The discovery of these small communities show that, rational randomness causes ordered diversity.

These small communities are also discovered by some of the community detection methods. However, former works do not pay any attention to them but give different kinds of appended manmade rule to achieve the fission results\cite{16}. In this paper, we try to deduce human invasion in community detection and recur the scene of real diverse community world which is the factual property of complex systems. With this idea, we get more instructional information.

\item Diverse core-community results in table.2 show that the individuals live a diverse life at normal state before the fission that the instructor node 1 left and opened another club. Every possible structure in Fig.1 appears with almost equal probability. Diverse real-community results in table.2 show that, the probability to divide into two parts is $56.34\%$, the probability to divide into 3 parts is $37.33\%$, the probability to divide into 4 parts is $6.33\%$. The structure with 2 communities appears apparently more frequently.

\item These results could also provide valuable information for the control of the dynamics of the evolution of communities in complex system. For example, it is explicitly easier, in the karate system, for the instructor node 1 to persuade nodes (25,26) to win the fission, and nodes (6,7,17) to persuade for the administer node 34 to win correspondingly.
\end{itemize}

The normal diverse state reveals that Zachary's karate club is a diverse real life system with major and minor communities. Diverse communities emerge from rational local interaction of individuals according to their preference as well as have no influence on the fission. That is, the karate system is a diverse complex system with hidden inside order.

We also tested our method with the dolphin system (38 components) and ISI social science journal system (1575 components).

\begin{figure}[!htp]
\centering
\includegraphics[width=5.5cm,height=4.5cm]{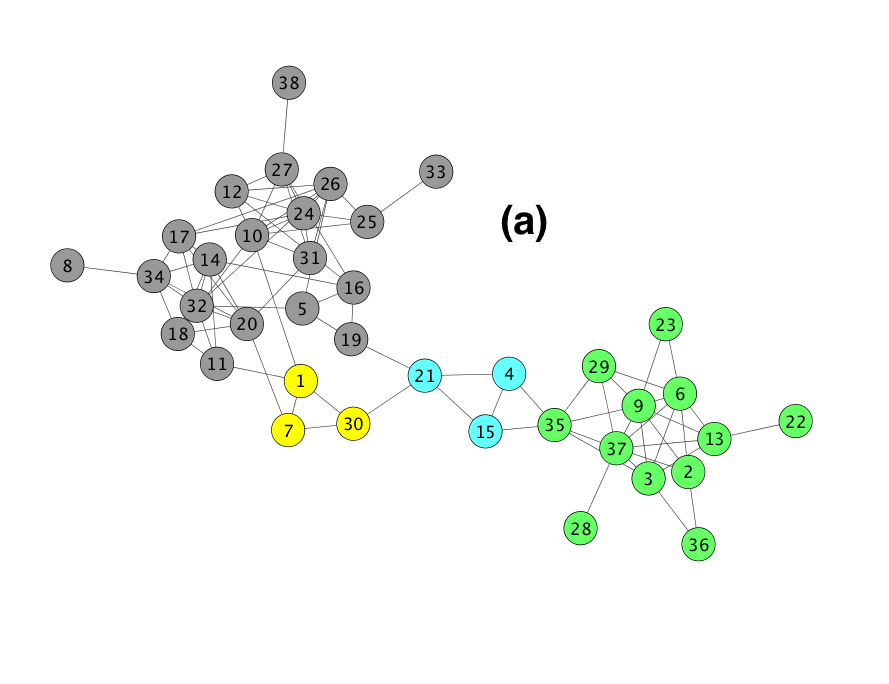}
\hspace{-0.25in}
\includegraphics[width=5.5cm,height=4.5cm]{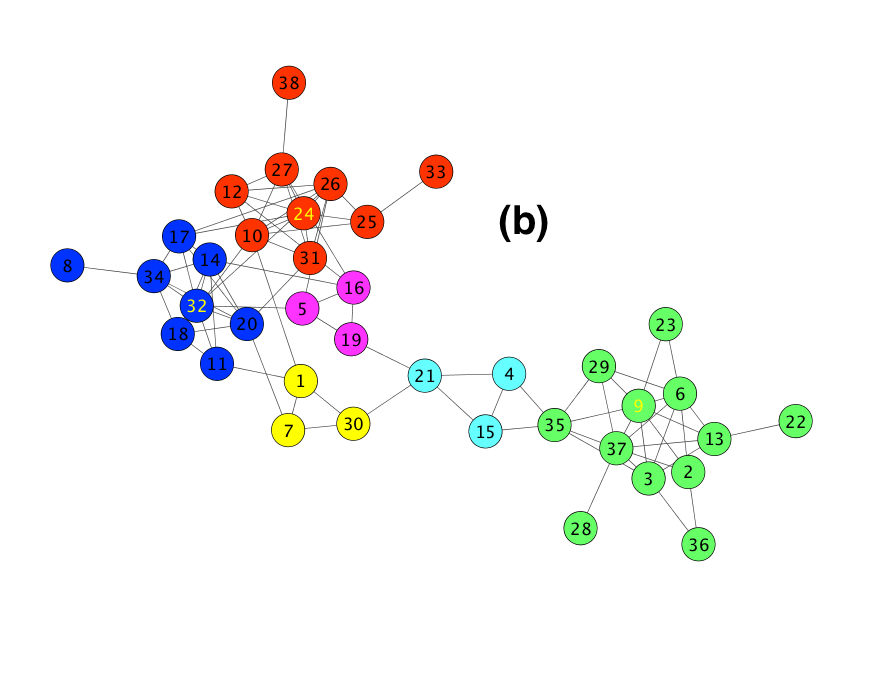}
%\hspace{0.2in}
\includegraphics[width=5cm,height=4cm]{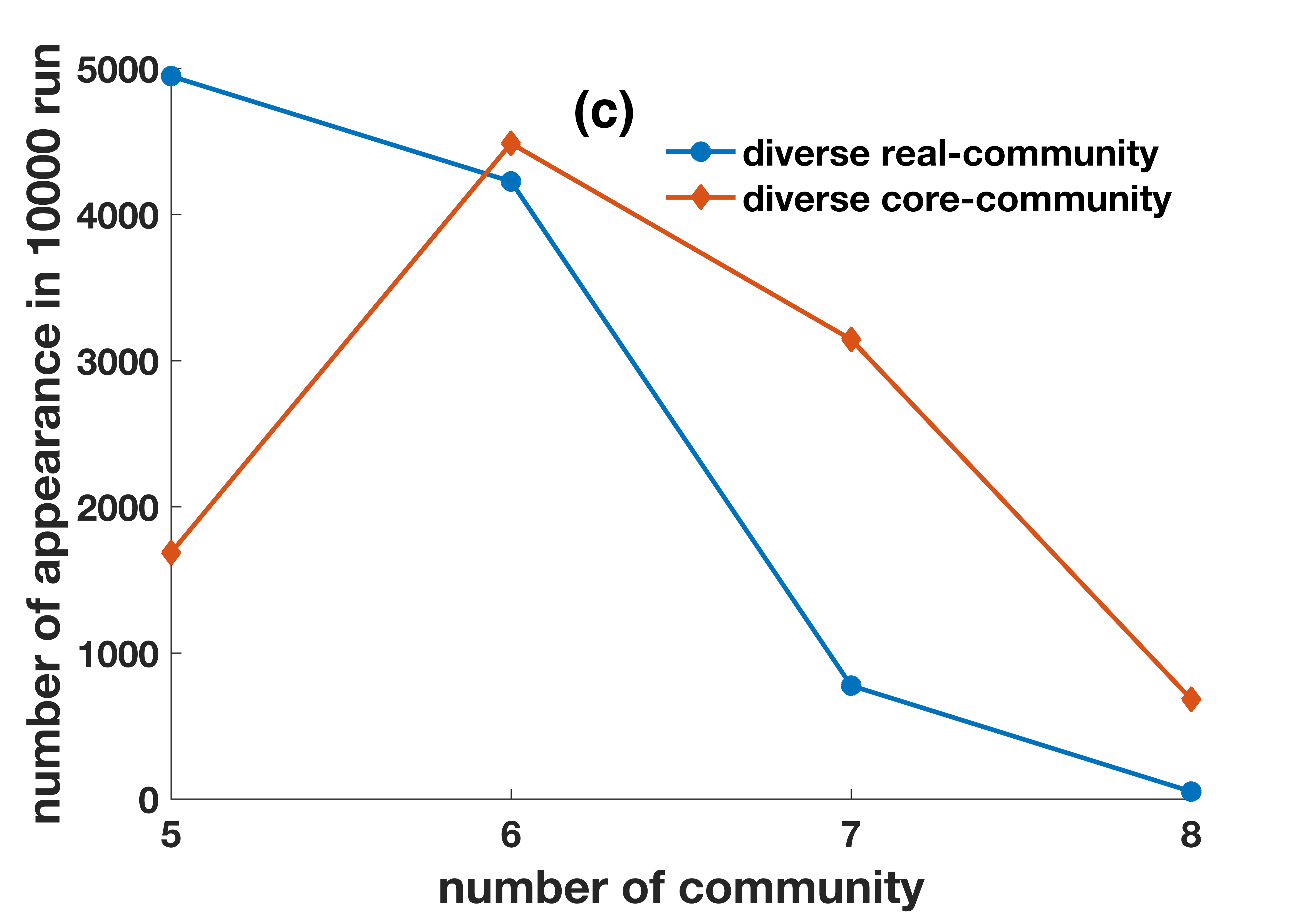}
\renewcommand{\figurename}{Figure}
\caption{Community results for the dolphin system.}
\end{figure}

In the dolphin system, the social network data contain 62 individuals\cite{31} while only 40 ones are included in their association analysis\cite{32}. As for these 40 dolphins, one (BZ) is not included in the 62 system, we could not get its connection information. And node (Natch) is only connected to BZ in the 40-system. So the remaining 38 dolphins are selected as our research objects. With the above community detection method, we found all the three associations and their core dolphins mentioned in \cite{32}. Figure 2 is the community detection results for the dolphin system. Figure 2(a) is the 4 real-communities and Figure 2(b) is the 6 core-communities in the deterministic hidden state. The three major communities in Figure 2(b) correspond to the three significant associations found by D. Lusseau\cite{32}, with nodes 9 (Gallatin), 24 (Scabs) and 32(Topless) hold the central position correspondingly. The communities leaded by 24 (red) and 32 (blue) are the two significant male associations, and the community leaded by 9 (green) is the female association. These results are surprisingly in accordance with the observation results in reference\cite{32}.

Figure 2(c) is the distribution of the number of community in the normal diverse state. The distribution indicates that the dolphins live a diverse life with more communities at normal state. These diversities occur in the biggest community in Figure 2(a) (grey, composed by the two male associations), while the female community remains unchanged all the time. This stability indicates that the dolphins live in a diverse social association with hidden inside order.

\begin{table}[!htp]
\begin{center}
\begin{tabular}{|c | c | c |}
\hline
state & type of community & number of communities\\
\hline
hiddern & RCom &1 \\ 
 \cline{2-3}
 deterministic & CCom & 11\\
 \hline
normal & RCom & 3 \\
\cline{2-3}
diversel & CCom & 10,11,12,13\\ 
\hline
\end{tabular}
\renewcommand{\thetable}{\arabic{table}}
\renewcommand{\tablename}{Table}
\caption{Number of communities for the ISI journal system.}
\end{center}
\label{default}
\end{table}%

\begin{figure}[!htp]
\centering
\includegraphics[width=5cm,height=4cm]{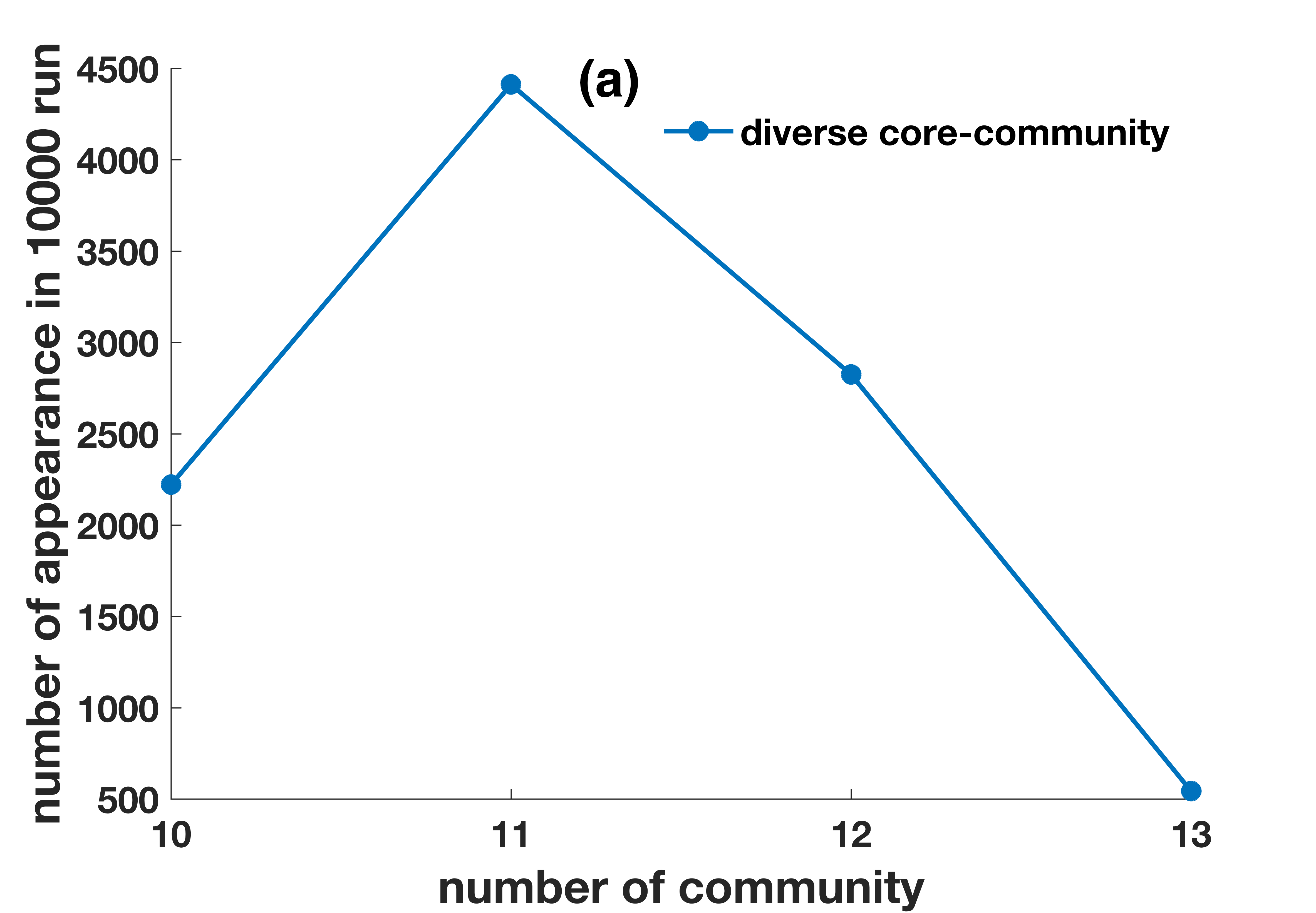}
\includegraphics[width=5cm,height=4cm]{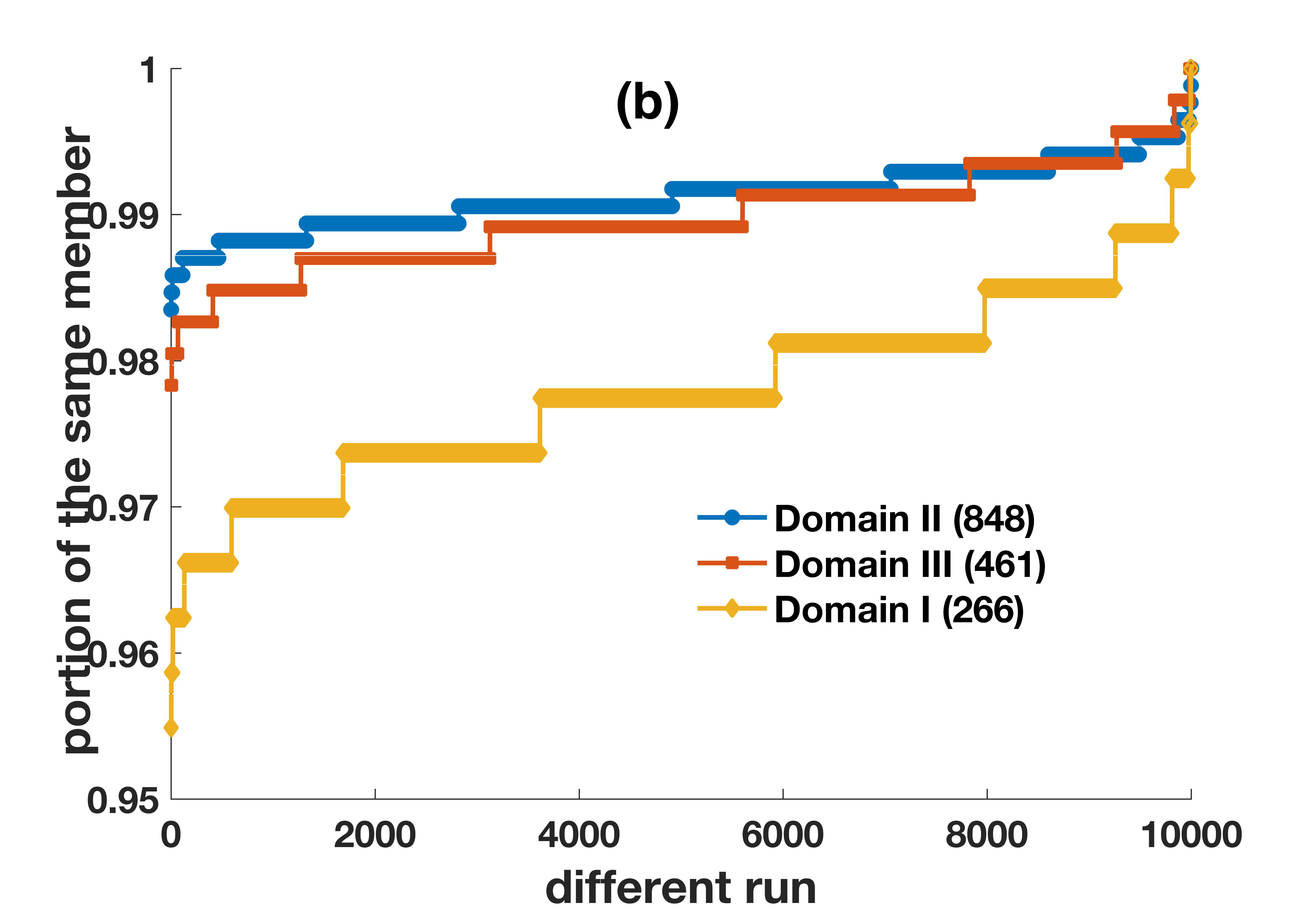}
\includegraphics[width=5cm,height=4cm]{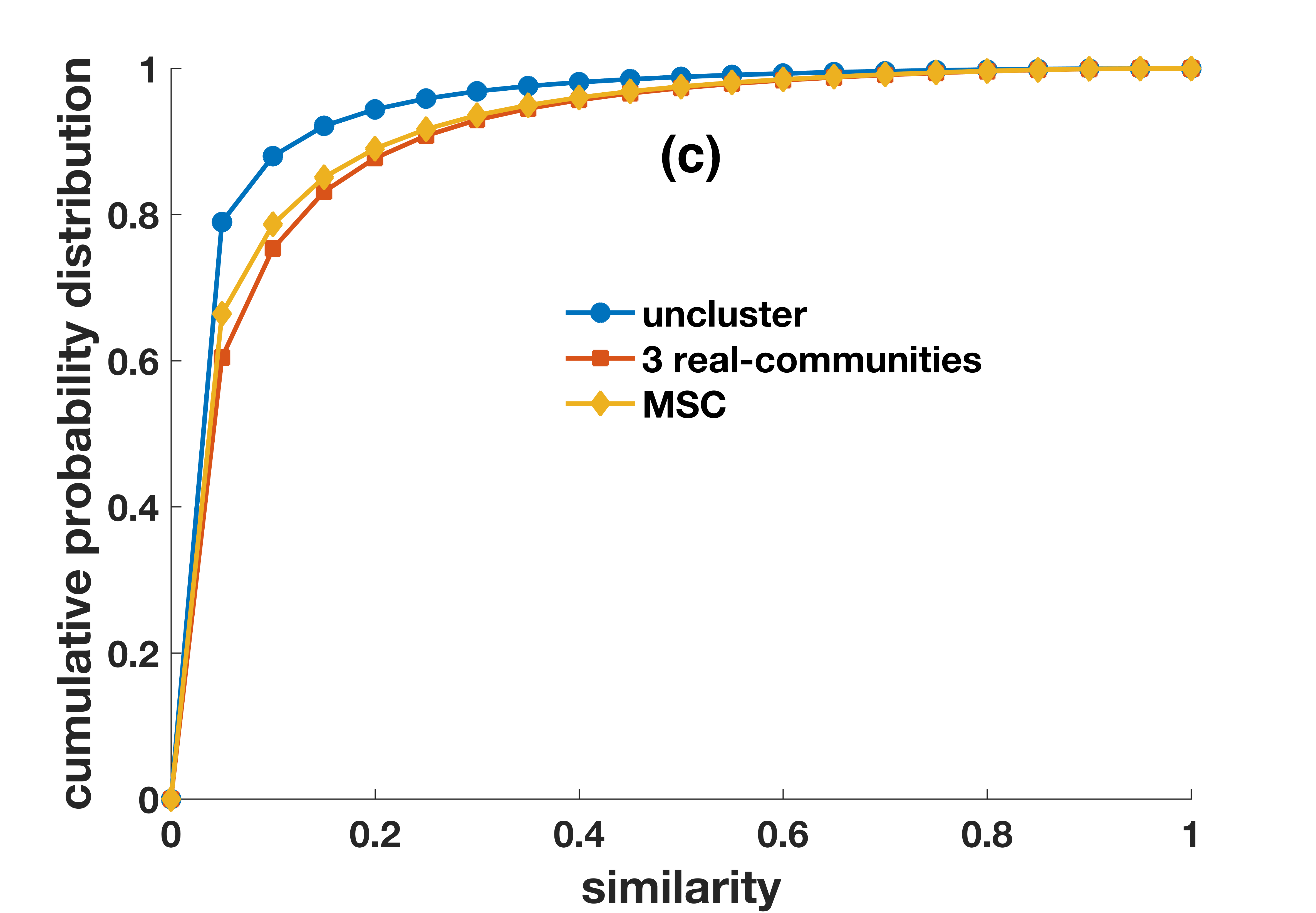}
\renewcommand{\figurename}{Figure}
\caption{Statistics of the community results for the ISI journal system.}
\end{figure}

Finally, we tested our method on a large complex system with 1575 nodes, the ISI social science journal system\cite{24}. In 2011, we used the cosine of co-citation as similarity, which could reflect the citation pattern of journals in detail. With this accurate information, every journal has only one most similar journal. That means, the similarity of co-citation pattern results in the same hidden deterministic and normal diverse community world. In this paper, we use connected common neighbor as similarity for comparison and simplicity. Table 3 is the community detection results. In the hidden deterministic state, these journals are all social science journals with 11 different research fields. And in the normal diverse state, three major research domains appear with 11.1689 different research fields on average. Figure 3(a) is the distribution of the number of communities in diverse normal state. These results are in accordance with our former work\cite{24}. Three real-communities in the normal diverse state correspond to the three knowledge domains. Domain I is the study of sociology includes sociology, politics and Geography with American sociological review (Impact Factor: 3.989) as the leading journal. Domain II is Psychology and related research field with Psychological Review (Impact Factor: 10.872) as the leading journal, this domain contains about half of the 1575 journals. Domain III is the study of social phenomenon includes History, Economics, Finance and Law, with American Historical Review (Impact Factor: 1.618) as the leading journal. These three leading journals are the most outstanding journals in its research domain correspondingly.

In order to show the hidden inside order of the normal diverse state, we made a comparison of the  3 real-communities in 10000 run. By setting the result of the first run (Domain I(266 journals), Domain II (848 journals), Domain II (461 journals)) as a  standard result for comparison, figure 3(b) gives out the portion of consistence of the member journals in sorted order. W can get from figure 3(b) that SSCI journals is a stationary system with 3 domains because more than $95.49\%$ of the journals keep their research domain unchanged during the 10000 run. Detailedly, for the 1575 journals, 282 of them have more than one most similar node pair (282 nodes with 402 most similar node pairs). However, of these 282 journals, 36 nodes are found lies on the boundary of different domain.

Figure 3(c) is the cumulative probability distribution of the journal journal similarity with different community detection methods. The curve labeled unclustered is the cumulative probability distribution for the original SSCI journal co-citation similarity. The curve labeled with MSC is the cumulative probability distribution with 3 research domains in our former work\cite{24}. The curve labeled diverse 3 core-communities is the cumulative probability distribution of this paper. Comparisons of the distribution show the efficiency of our method, which can detect the components that are similar with each other into the same community but not the dissimilar ones. It behaves even better than the former coarse-grained results.

{\bf Summary}

Communities are common and play a significant role in the functioning of complex systems. Inspired by the ultimatum game experiment in reference \cite{23} and our former work\cite{24}, in this paper, we propose one simple and efficient community detection method. Our method could detect the community structure of complex systems efficiently by merging components of the closest proximity. By introducing rational random selection, our method reveals the hidden deterministic and normal diverse community states of community structure. These two states have direct corresponding meaning in real-life application.  Most of the time, individuals live in a diverse world with many small communities until there happens some big event. We have many examples of this kind, just like the small research groups with same research interest in different country, small groups of Marathon all over the earth, human society before or after the election, etc. The community structure in different states correspond to various pint of view for viewing the community structure of real-life complex systems with diversity. We test our method with 3 real-world systems and find that the method could not only detect community structure with high sensitivity and reliability but could also provide instructional information about our community world by giving out the core-community, the real-community, the tide (boundary) and the diversity. 

What' more, statistics of the core-community  and real-community reveal the hidden inside properties of complex systems. These results also give a possible indication that it is the rational randomness based on self expectation that is significant in the emergence of diversity and stability of communities. 

{\bf Acknowledgements}

This work is supported by National Natural Science Foundation of China (Grant No. 11547003), and China Scholarship Council (Grant No. 201607620007). 

{\bf Author contributions statement}

Y.F.C. conceived the method, performed the experiment, analyzed the results and prepared the manuscript. S.K.H. gave helpful comments on the analysis of the results and the manuscript. X.D.W. gave helpful comments on the manuscript.

%This work is supported by National Natural Science Foundation of China (Grant No. 11547003), and China Scholarship Council (Grant No. 201607620007). \par

%=================================================================================================

%======================================================================================================

\begin{thebibliography}{999}
%spintronic devices

\bibitem{1} G.W. Flake, S. Lawrence, C.L. Giles  and F.M. Coetzee, Self-organization and identification of Web communities, IEEE Computer, \textbf{35(3)}, 66-71, (2002).
\bibitem{10} M. Girvan and M.E.J. Newman, Community structure in social and biological networks, Proceedings of the National Academy of Sciences, \textbf{99(12)}, 7821-7826, (2002).
\bibitem{2} E. Ravasz, A.L. Somera, A. Mongru, Z.N. Oltvai, A.L. Barabasi, Hierarchical organization of modularity in metabolic networks, Science, \textbf{297(5586)}, 1551-1555, (2002).
\bibitem{3} R. Guimer\`{a} and L.A.N. Amaral, Functional cartography of complex metabolic networks, Nature, \textbf{433}, 895-900 (2005).
\bibitem{4} G. Palla, I. Der\'{e}nyi, I. Farkas and T. Vicsek, Uncovering the overlapping community structure of complex networks in nature and society, Nature, \textbf{435}, 814-818, (2005).
\bibitem{5} M. Huss and P. Holme, Currency and commodity metabolites: Their identification and relation to the modularity of metabolic networks, IET Systems Biology, \textbf{1(5)}, 280-285, (2007).
%ratchet effect

\bibitem{6} M.C. Gonzalez, C.A. Hidalgo and A.L. Barabasi, Understanding individual human mobility patterns, Nature, \textbf{453}, 779-782, (2008).
\bibitem{7} A.C. Gavin, et al., Proteome survey reveals modularity of the yeast cell machinery, Nature, \textbf{440}, 631-636, (2006).
\bibitem{8} J. Kleinberg and S. Lawrence, The structure of the Web, Science, \textbf{294}, 1849-1850, (2001).
\bibitem{9} S.N. Dorogovtsev and J.F.F. Mendes, Evolution of Networks: From Biological Nets to the Internet and WWWW, Oxford University Press, (2003).
\bibitem{11} S. Boccaletti, V. Latora, Y. Moreno, M. Chavez, D.U. Hwang, Complex networks: structure and dynamics, Physics Report, \textbf{424(4-5)}, 175-308, (2006). 
%10 Girvan M, Newman MEJ. Community structure in social and biological networks. Proceedings of the National Academy of Sciences. 2002;99(12):7821?7826.
\bibitem{42} L. Danon, A. D\'{i}az-Guilera, J. Duch and A. Arenas, Comparing community structure identification, Journal of Statistical Mechanics: Theory and Experiment, \textbf{09}, P09008, (2005).
\bibitem{43} S. Fortunato, Community detection in graphs, Physics Reports, \textbf{486(3?5)}, 75-174, (2010).

\bibitem{33} J.P. Bagrow and E.M. Bollt, Local method for detecting communities. Physical Review E, \textbf{72}, 046108, (2005).
\bibitem{34} M.E.J. Newman, Modularity and community structure in networks. Proceedings of the National Academy of Sciences, \textbf{103(23)}, 8577-8582, (2006).
\bibitem{35} V.D. Blondel, J.L. Guillaume, R. Lambiotte and E. Lefebvre, Fast unfolding of communities in large networks, Journal of Statistical Mechanics: Theory and Experiment, \textbf{10}, P10008, (2008).
\bibitem{36} U.N. Raghavan, R. Albert and S. Kumara, Near linear time algorithm to detect community structures in large-scale networks, Physical Review E, \textbf{76}, 036106, (2007).
\bibitem{37} S. Gregory, Finding overlapping communities in networks by label propagation, New Journal of Physics, \textbf{12(10)}, 103018, (2010).
\bibitem{38} J. Xie and B.K. Szymanski, LabelRank: A stabilized label propagation algorithm for community detection in networks, In Network Science Workshop (NSW), 2013 IEEE 2nd, 138?143, (2013).
\bibitem{39} B. Karrer and M.E.J. Newman, Stochastic blockmodels and community structure in networks, Physical Review E, \textbf{83}, 016107, (2011).
\bibitem{40} R. Aldecoa and I. Mar\'{i}n, Surprise maximization reveals the community structure of complex networks. Scientific Reports, \textbf{3}, 1060, (2013).
\bibitem{41} A. Lancichinetti, F. Radicchi, J.J. Ramasco and S. Fortunato, Finding Statistically Significant Communities in Networks. PLoS ONE, \textbf{6(4)}, 1-18, (2011).

%\bibitem{12} M.E.J. Newman and M. Girvan, Finding and evaluating community structure in networks, Physics Review E, \textbf{69}, 026113, (2004). 
%\bibitem{13} A. Pothen, H. Sinmon and K.-P. Liou, Partitioning sparse matrices with eigenvectors of graphs, SIAM Journal on Matrix Analysis and Applications, \textbf{1(11)}, 430-452, (1990).
%\bibitem{14} C. Shi, Z.Y. Yan, Y. Wang, Y.N. Cai and B. Wu, A genetic algorithm for detecting communities in large-scale complex networks, Advance in Complex System, \textbf{13(1)}, 3-17, (2010).
%\bibitem{15} U.N. Raghavan, R. Albert and S. Kumara, Near linear time algorithm to detect community structures in large-scale networks, Physics Review E, \textbf{76}, 036106, (2007).

\bibitem{17} G. P. Garnett, J. P. Hughes, R. M. Anderson, B. P. Stoner, S. O. Aral, W. L. Whittington, H. H. Handsfield, and K. K. Holmes, Sexual mixing patterns of patients attending sexually transmitted diseases clinics, Sexually Transmitted Diseases, \textbf{23}, 248-257, (1996).
\bibitem{18} S. O. Aral, J. P. Hughes, B. Stoner, W. Whittington, H. H. Handsfield, R. M. Anderson, and K. K. Holmes, Sexual mixing patterns in the spread of gonococcal and chlamydial infections. American Journal of Public Health, \textbf{89}, 825-833, (1999).

\bibitem{26} M. E. J. Newman, Detecting community structure in networks. Eur. Phys. J. B, \textbf{38}, 321-330, (2004).
\bibitem{27} M. Gustafsson, A. Lombardi, and M. Hornquist, Comparison and validation of community structures in complex networks, Physica A, \textbf{367}, 559-576, (2006).

\bibitem{19} O. Sporns, D.R. Chialvo, M. Kaiser, and C.C. Hilgetag, Organization, development and function of complex brain networks. Trends in Cognitive Sciences, \textbf{8(9)}, 418-425,  (2004).

\bibitem{20} B.J. Frey, and D. Dueck, Clustering by passing messages between data points. Science, \textbf{315(5814)}, 972-976, (2007). 
\bibitem{21} C.D. Manning, P. Raghavan and H. Sch\"{u}tze, Introduction to information retrieval. Cambridge, UK: Cambridge University Press, (2008).
\bibitem{22} T. Hastie, R. Tibshirani and J. Friedman, The Elements of Statistical Learning (2nd ed.). New York: Springer-Verlag, (2008). 
\bibitem{23} X. Han, S.N. Cao, Z.S. Shen, B.Y. Zhang, W.X. Wang, R. Cressman and H.E. Stanley, Emergence of communities and diversity in social networks, Proceedings of the National Academy of Sciences, \textbf{114(11)}, 2887-2891, (2017).
\bibitem{24} Y.F. Chang and C.M. Chen, Classification and visualization of the social science network by the minimum span clustering method, Journal of the American Society for Information Science and Technology, \textbf{62(12)}, 2404-2413, (2011).
\bibitem{25} W. W. Zachary, An information flow model for conflict and fission in small groups, Journal of Anthropological Research, \textbf{33}, 452-473, (1977).
\bibitem{28} Y. Dourisboure, F. Geraci and M. Pellegrini,  Extraction and classification of dense communities in the web, In Proceedings of the 16th international conference on World Wide Web, pp, 461-470, (2007)
\bibitem{29} E. Ostrom, Understanding Institutional Diversity, Princeton University Press, (2009). 
\bibitem{30} A.W. Rives, T. Galitski, Modular organization of cellular networks, Proceedings of the National Academy of Sciences, \textbf{100(3)}, 1128-1133, (2003). 

\bibitem{16} K.R. Z\v{a}lik, Maximal neighbor similarity reveals real communities in networks, Scientific Reports, \textbf{5}, 18374, (2015).
\bibitem{31} D. Lusseau, The emergent properties of a dolphin social network, Proc. R. Soc. London B (suppl.), \textbf{270}, S186-S188 (2003).
\bibitem{32} D. Lusseau, K. Schneider, O.J. Boisseau, P. Haase, E. Slooten and S.M. Dawson, The bottlenose dolphine community of doubtful sound features, Behav. Ecol. Sociobiol., \textbf{54}, 396-405, (2003).


\end{thebibliography}
\end{document}